\documentstyle[aps,epsf]{revtex}

\twocolumn

\begin{document}
% Final version: two column
%\noindent File {\sf PILIGO.tex}\\ \hfill {Begined may 29, 2002} \hfill {Printed \today}
\title{Analysis of Parametric Oscillatory Instability\\
     in Power Recycled LIGO Interferometer}
\author{
V. B. Braginsky, S. E. Strigin and S. P. Vyatchanin}
\address{
Faculty of Physics , Moscow State University, Moscow 119992,
	Russia 
	\\  e-mail: vyat@hbar.phys.msu.su}
\date{\today}
\maketitle
\begin{abstract}
We present the analysis of a nonlinear effect of parametric
oscillatory instability in power recycled LIGO interferometer with
the Fabry-Perot (FP) cavities in the arms. The basis for this
effect is the excitation of the additional (Stokes) optical mode
with frequency $\omega_1$ and the mirror elastic mode with
frequency $\omega_m$, when the optical energy stored in the main
FP cavity mode with frequency $\omega_0$ exceeds the certain
threshold and the frequencies are related as $\omega_0\simeq
\omega_1+\omega_m$. The presence  of anti-Stokes modes (with
frequency $\omega_{1a}\simeq \omega_0+\omega_m$) can  depress
parametric instability. However, it is very likely that the
anti-Stokes modes will not compensate the parametric instability
completely.

\end{abstract}

\section{Introduction}

The full scale terrestrial interferometric gravitational wave
antennae are in process of assembling and tuning at present. One
of these antennae (LIGO-I project) sensitivity expressed in terms
of the metric perturbation amplitude is projected to achieve the
level of $h\simeq 1\times 10^{-21}$ \cite{abr1,abr2}. After the
improvement of the isolation from noises in test masses (the
mirrors of the 4 km long optical FP cavities) and after increasing
the optical power circulating in the resonator up to $W\simeq
830$~kW the sensitivity is expected to reach the value of $h\simeq
1\times 10^{-22}$ \cite{amaldi}. This value of $W$ corresponds to
the energy ${\cal E}_0 \simeq 22$~J stored in the FP resonator.

In previous paper \cite{bsv} we have described the possibly
existing effect of  pure nonlinear dynamical origin which may
cause substantial decrease of the antennae sensitivity or even the
antenna disfunction. The essence of this effect is classical
parametric oscillatory instability in the FP cavity which modes
are coupled with mechanical degree of freedom of the mirror. This
effect appears above the certain threshold of the optical energy
${\cal E}_0$ when the difference $\omega_0 -\omega_1$ between the
frequency $\omega_0$ of the main optical mode (which stores ${\cal
E}_0$) and the frequency $\omega_1$ of the idle (Stokes) mode is
close to the frequency $\omega_m$ of the mirror mechanical degree
of freedom. The coupling between these three modes appears due to
the ponderomotive pressure of the light photons in main and 
Stokes modes and due to the parametric action of mechanical
oscillation on the optical modes. Above the critical value of
energy ${\cal E}_0$ (dimensionless parameter ${\cal R}_0 >1$, see
below) the amplitude of mechanical oscillation will rise
exponentially as well as the optical power in the idle (Stokes)
optical mode.

In the article \cite{bsv} we have used a simplified model of this
effect in which lumped model of mechanical oscillator had been
used and optical modes with gaussian distribution over the cross
section had been taken into account. Under these assumptions the
parametric oscillatory instability will appear if
\begin{eqnarray}
\label{CI}
\frac{{\cal R}_0}{\left(1+\frac{\Delta \omega_1^2}{\delta_1^2}\right) } & > & 1,\\
\label{R0}
{\cal R}_0  =  \frac{{\cal E}_0 }{2mL^2\omega_m^2}\,
    \frac{\omega_1\omega_m}{\delta_1\delta_m}&=&
    \frac{{2\cal E}_0 Q_1Q_m}{mL^2\omega_m^2}\,.
\end{eqnarray}
Here $m$ is the value of the order of mirror mass, $L$ is the
distance between the FP cavity mirrors, $\Delta
\omega_1=\omega_0-\omega_1 -\omega_m$, $\delta_1$ and
$Q_1=\omega_1/2\delta_1$ are the relaxation rate and quality factor of
the Stokes mode correspondingly, $\delta_m$ and
$Q_m=\omega_m/2\delta_m$ are relaxation rate and quality factor of
the mechanical oscillator.

%For estimates the following numerical parameters were assumed:
%\begin{equation}
%\label{paramold}
%\begin{array}{lcllcl}
%\omega_m &=& 2\times 10^5 \ {\rm sec}^{-1},&
%    \delta_m &=&5\times 10^{-3} \ {\rm sec}^{-1},\\
%\delta_1 &=& 6\times 10^2\ {\rm sec}^{-1}, &
%    \omega_1& \simeq & 2\times 10^{15}\ {\rm sec}^{-1},\\
%{\cal E}_0 &\simeq& 3\times 10^8\ {\rm erg},& L&=&4\times 10^5\ {\rm cm}\\
%m &=& 10^4\ {\rm g},&  &&
%\end{array}
%\end{equation}

Recently E. D'Ambrosio and W. Kells \cite{ak} have reported that
if in the same one dimensional model the anti-Stokes  mode (with
frequency $\omega_{1\, a}= \omega_0 +\omega_m$)  is taken into
account then the effect of parametric instability will be
substantially dumped or even excluded. In this article we  present
the analysis based on the model which takes into account several
important details of the antenna. This analysis shows that
parametric oscillatory  instability still may exist.

In section \ref{sec1}
we present the analysis of one dimensional optical
model of antenna in which the so called power recycled mirror is
taken into account. In section \ref{sec2}
important key elements of 3-dimensional
approach are used to prove that it is very likely
that the anti-Stokes modes will not compensate the oscillatory instability.

\section{ The role of power recycling mirror in the antenna}\label{sec1}

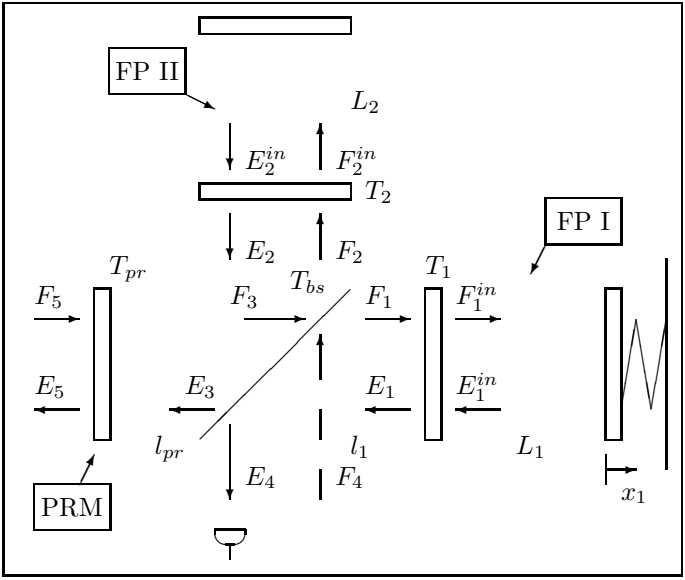
\begin{figure}
\noindent
\unitlength0.1mm
\linethickness{0.5pt}
\begin{picture}(914,762)
\put(120,180){\framebox(20,200)[]{}}
\put(40,340){\vector(1,0){60}}
\put(320,340){\vector(1,0){80}}
\put(560,180){\framebox(20,200)[]{}}
\put(800,180){\framebox(20,200)[]{}}
\put(480,340){\vector(1,0){60}}
\put(260,500){\framebox(200,20)[]{}}
\put(260,720){\framebox(200,20)[]{}}
\put(300,480){\vector(0,-1){60}}
\put(300,200){\vector(0,-1){100}}
\put(480,360){$F_1$}
\put(600,340){\vector(1,0){60}}
\put(300,600){\vector(0,-1){60}}
\put(540,220){\vector(-1,0){60}}
\put(100,220){\vector(-1,0){60}}
\put(480,240){$E_1$}
\put(660,220){\vector(-1,0){60}}
\put(420,420){\vector(0,1){60}}
\put(440,420){$F_2$}
\put(420,540){\vector(0,1){60}}
\put(320,420){$E_2$}
\put(560,400){$T_1$}
\put(480,500){$T_2$}
\put(420,180){\line(0,1){40}}
\put(420,260){\vector(0,1){60}}
\put(680,160){$L_1$}
\put(460,620){$L_2$}
\put(820,220){\line(1,6){20}}
\put(840,340){\line(1,-6){20}}
\put(860,220){\line(1,6){20}}
\put(880,420){\line(0,-1){280}}
\put(0,0){\framebox(900,760)[]{}}
\put(300,60){\oval(40,40)[b]}
\put(280,60){\line(1,0){40}}
\put(600,240){$E_1^{in}$}
\put(320,540){$E_2^{in}$}
\put(300,40){\line(0,-1){20}}
\put(440,540){$F_2^{in}$}
\put(600,360){$F_1^{in}$}
\put(380,380){$T_{bs}$}
\put(300,360){$F_3$}
\put(460,380){\line(-1,-1){200}}
\put(460,160){$l_1$}
\put(200,160){$l_{pr}$}
\put(280,220){\vector(-1,0){60}}
\put(240,240){$E_3$}
\put(420,140){\line(0,-1){40}}
\put(140,400){$T_{pr}$}
\put(320,120){$E_4$}
\put(40,360){$F_5$}
\put(40,240){$E_5$}
\put(440,120){$F_4$}
\put(720,440){\framebox(100,60)[]{FP I}}
\put(720,440){\vector(-1,-2){20}}
\put(140,640){\framebox(100,60)[]{FP II}}
\put(240,640){\vector(2,-1){40}}
\put(800,160){\line(0,-1){40}}
\put(800,140){\vector(1,0){40}}
\put(820,100){$x_1$}
\put(40,60){\framebox(100,60)[]{PRM}}
\put(100,120){\vector(1,2){20}}
\end{picture}\\[2mm]
\caption{The scheme of power recycled LIGO  interferometer
    with the Fabry-Perot (FP) cavities in the arms. The end
    mirror of the FP I cavity is a part of mechanical oscillator.
    $T_{bs}$ and $T_{pr}$ are transmittances of the beam splitter
    and power recycling mirror correspondingly, $l_1$ is the
    distance between the input mirror of the FP I cavity and the beam splitter,
    $l_{pr}$ is the distance between the power recycling mirror and
    the beam splitter, $F_i$ and $E_i$ are the complex amplitudes of
    travelling optical waves in different parts of interferometer.}\label{fig1}
\end{figure}

The design  of laser interferometer gravitational wave antenna
apart from the two main optical FP cavities also includes the so
called power recycling mirror (PRM) which allows to increase the
value of ${\cal E}_0$ using the same laser input power (see
fig.~\ref{fig1}).

The results of calculations presented below are based on
the following simplifying assumptions:
\begin{itemize}
\item   Only one mirror (in the FP I cavity  in fig.~\ref{fig1}) is movable,
        it is a part of mechanical oscillator that is considered as
        lumped one with single mechanical degree of freedom (eigenfrequency
        $\omega_m$,  quality factor $Q_m=\omega_m /2\delta_m$ and mass $m$
        which is of the order of the total mirror mass). All other
        mirrors are assumed to be fixed.
\item   All mirrors have no optical losses. Both end mirrors of FP cavities
        have the ideal reflectivity.
        The input mirrors of the FP cavities are identical and have the
        finite transmittances $T_1=T_2=T=2\pi L/(\lambda_0 Q_{opt})$
        ($\lambda_0$ is the optical wavelength,
        $Q_{opt}$ is the quality factor, $L$ is the distance between the
        mirrors). The distances $L_1=L_2=L$ are equal.
\item   Only the main (pumped) mode with frequency  $\omega_0$
        and relaxation  rate $\delta_0=\omega_0/2Q_0$,  and
        Stokes mode with $\omega_1$ and $\delta_1=\omega_1/2Q_1$
        correspondingly are taken into account
        ($Q_0$ and $Q_1$ are the quality factors). It is assumed that
        $\omega_0-\omega_1 \simeq \omega_m$.
\item   Only the main mode is pumped by laser and the value of
        stored energy ${\cal E}_0$
        is constant (approximation of
        constant field).
\item   In this particular model we do not take into account the
        possible influence of the anti-Stokes mode.
\end{itemize}

It is possible to calculate at what level of energy ${\cal E}_0$
the Stokes mode and mechanical oscillator become unstable. The
origin of this instability can be described qualitatively in the
following way: small mechanical oscillations with resonance
frequency $\omega_m$ modulate the distance $L$ that causes the
excitation of optical fields with frequencies $\omega_0 \pm
\omega_m$. Therefore, the Stokes mode amplitude will rise linearly
in time if time interval is shorter than relaxation time. The
presence of two optical fields with frequencies $\omega_0$ and
$\omega_1$ will produce the component of ponderomotive force
(which is proportional to the square of fields sum) at difference
frequency $\omega_0-\omega_1$. Thus this force will increase the
initially small amplitude of mechanical oscillations.
%It is worth to note that in power recycled interferometer Stokes wave
%irradiated from input mirror of FP cavity will partially returned back due
%to power recycling mirror --- therefore, the parametric instability effect in this interferometer
%must be larger
%than in separate FP cavity.

For analysis of parametric instability we have to use two
equations for the Stokes mode and mechanical oscillator, and find
the conditions when this "feedback" prevails the damping which
exists due to the finite values of relaxation in mechanical
resonator and in the FP cavity. Note that in resonance
$\omega_0\simeq \omega_1 +\omega_m$ the effect of parametric
instability for power recycled LIGO interferometer is larger than
for the separate FP cavity because the Stokes wave (at frequency
$\omega_1$) emitted from the FP cavity   throughout its input
mirror is not lost irreversible but returns back due to power
recycling mirror, therefore, its interaction is prolonged.

We can write down the field components $E_0^{in}$, $E_1^{in}$ of the
main and Stokes modes inside the FP I cavity correspondingly and
the displacement $x$ of mechanical oscillator in rotating wave
approximation as:
\begin{eqnarray}
E_0^{in} &=& A_0[D_0e^{-i\omega_0 t}+D_0^*e^{i\omega_0 t}],\nonumber\\
\label{sa}
E_1^{in} &=& A_1[D_1e^{-i\omega_1 t}+D_1^*e^{i\omega_1 t}], \\
x &=& Xe^{-i\omega_m t} + X^* e^{i\omega_m t},\nonumber
\end{eqnarray}
where $D_0$ and $D_1$ are the slowly changing complex amplitudes
of the main and Stokes modes correspondingly, and $X$ is the
slowly changing complex amplitude of mechanical displacement.
Normalizing constants $A_0,\ A_1$ are chosen so that energies
${\cal E}_{0,\,1}$ stored in each mode (of the FP I cavity) are
equal to ${\cal E}_{0,\,1}=\omega_{0,\, 1}^2|D_{0,\, 1}|^2/2$.
Then one can obtain the equations for slowly changing amplitudes
(see details in Appendix \ref{app1}):
\begin{eqnarray}
\label{D1a}
\left(\partial_t + \delta_1 \right)(\partial_t+\delta_{pr})\, D_1^*
&=& \frac{-i D_0^*\omega_0}{ L}\times\\
\times  \left[\partial_t+\delta_{pr} +\frac{\delta_1}{2}\right]&\cdot &
        X\, e^{i\Delta \omega_1 t} , \nonumber \\
\Delta\omega_1 & = & \omega_0-\omega_1-\omega_m,\nonumber\\
\delta_1=\frac{cT}{4L},\quad \delta_{pr} &=& \frac{T_{pr}\delta_1}{4}\ll \delta_1,\nonumber\\
\label{Xb}
\left(\partial_t + \delta_m \right)\, X &=&
    \frac{iD_0D_1^*\omega_0\omega_1}{m\omega_m L} e^{-i\Delta \omega_1
    t}\,.
\end{eqnarray}
Remind that we assume $D_0$ to be constant. The additional
relaxation rate $\delta_{pr}$ describes the relaxation of
oscillations in the FP cavity with power recycling mirror.

One can find the solutions of (\ref{D1a}, \ref{Xb}) in the following
form
\begin{eqnarray*}
D_1^*(t) &=&  D_{1}^*e^{\lambda_1 t},\quad
  \lambda_1= \lambda +i\Delta\omega_1\\
X(t) &=& X\, e^{\lambda t},
\end{eqnarray*} and write down the
characteristic equation:
\begin{eqnarray}
\label{shchar}
\left(\lambda + \delta_m \right) &=&
    \frac{ {\cal R}_0\delta_1\delta_m}{\lambda_{1}+\delta_1}
    \left[1 +
      \frac{\delta_1}{ 2(\lambda_{1}+\delta_{pr})}\right].\\
%\label{R0}{\cal R}_0 & = & \frac{{\cal E}_0 }{2mL^2\omega_m^2}\,    \frac{\omega_1\omega_m}{\delta_1\delta_m} =    \frac{{2\cal E}_0 Q_1Q_m}{mL^2\omega_m^2}
\end{eqnarray}
The parametric oscillatory instability will appear if one of the
characteristic equation roots real part is positive. Analysis of
this equation with assumption
$$
\delta_m\ll \delta_{pr}\ll \delta_1,
$$
gives the condition of parametric instability
(see details in Appendix \ref{app2}):
\begin{eqnarray}
\label{PICOND}
\frac{{\cal R}_0}{ \left (
  1+\frac{\Delta\omega_{1}^2}{\delta_1^2} \right) }
    \times
      \frac{2+\frac{\delta_1}{\delta_{pr}}+
        \frac{\Delta\omega_{1}^2}{\delta_{pr}^2} }{
          2\left(1+\frac{\Delta\omega_{1}^2}{\delta_{pr}^2} \right) } & >
        & 1\,.
\end{eqnarray}
In ultimate resonance case when $\Delta\omega_1 \ll \delta_{pr}$
we obtain the condition of
resonance parametric instability:
\begin{equation}
\label{RPicond}
{\cal R}_0\times \left(1 +\frac{\delta_1}{2\delta_{pr}}\right)> 1.
\end{equation}
It means that the parametric instability takes place at energy
${\cal E}_0$ which is smaller than one for the separate FP cavity by
the factor of $\sim \frac{\delta_1}{2\delta_{pr}}$.

It is important that factors ${\cal R}_0$ and
$\frac{\delta_1}{2\delta_{pr}}$ have large numerical values, using
the parameters for probe mass fabricated from high quality
\cite{ageev} fused silica that are planned to use in LIGO-II we
obtain (see \cite{ligo2} and Appendix \ref{param}):
\begin{equation}
{\cal R}_0 \simeq 6100,\quad \frac{\delta_1}{2\delta_{pr}}\simeq
31\,.
\end{equation}
It means that if $\Delta\omega_1\ll \delta_{pr}$ the maximal
energy stored in the FP cavity can not exceed the value of about
$6100\times 31\sim 1.9\times 10^{5}$ times {\em smaller} than one
planned for LIGO-II (!).

Note that the estimate of ${\cal R}_0$ differs from the estimate
presented in \cite{bsv} because here we use parameters  more close
to ones planned for LIGO-II. In particular: {\em (a)}  mass is
four times greater than in \cite{bsv}; {\em (b)} mechanical
frequency is assumed to be two times smaller; {\em (c)} loss angle
(and mechanical relaxation rate) is $\sim 4$ times smaller in
accordance with results of Ageev and Penn \cite{ageev}; {\em (d)}
relaxation rate $\delta_1$ is $\sim 6$ times smaller.

The probability of the ultimate resonance ($\Delta\omega_1\ll
\delta_{pr}$) is extremely low because the value of $\delta_{pr}
\simeq 1.5\ {\rm s}^{-1}$ is rather small. In more realistic case
when $\Delta\omega_1^2 \gg \delta_1\delta_{pr}$ we obtain the
"partial resonance" condition of parametric instability from
(\ref{PICOND}):
\begin{eqnarray}
\label{FPcond}
\frac{{\cal R}_0}{2 \left (
  1+\frac{\Delta\omega_{1}^2}{\delta_1^2} \right) }
     & > & 1\,.
\end{eqnarray}
It means that in the case when $\delta_1 \gg \Delta\omega_1\gg
\sqrt{\delta_1\delta_{pr} }$ the maximal energy stored in the FP
cavity can not exceed the value of about $3050$ times {\em
smaller} than one planned for LIGO-II (!). Note that  condition
(\ref{FPcond}) differs from the parametric instability condition
in the separate FP cavity \cite{bsv} by the factor of $2$ in
denominator.

The considered model is the simplest one and more detailed
3-dimensional model of interferometer has to be analyzed. In
particular, there are reasons to hope that the danger of
parametric instability may be smaller in real interferometer:
\begin{itemize}
\item Even in resonance the Stokes and elastic modes may not
    spatially suit  to each other (small overlapping factor).
\item The possible presence of the anti-Stokes mode may partially or
    completely depress  the parametric instability.
\end{itemize}

In the next section we consider both these factors.

\section{Consideration on Three-Dimensional Analysis}\label{sec2}

It is possible to generalize the above simplified model  for the
arbitrary elastic mode in the mirror. It has been shown \cite{bsv}
that in this case the constant ${\cal R}_0$ in condition
(\ref{PICOND}) should be multiplied by the overlapping factor
$\Lambda_1$.

In general case when both Stokes and anti-Stokes modes have to be
taken into account the characteristic equation may be presented in
the following form:
\begin{eqnarray}
\label{shchar2}
\left(\lambda + \delta_m \right) &=&{\cal R}_0\delta_1\delta_m\times
    \frac{\Lambda_{1}}{\lambda_{1}+\delta_1}
    \left[1 +
      \frac{\delta_1}{ 2(\lambda_{1}+\delta_{pr})}\right]-\nonumber\\
\label{chareq2AS}
&&- {\cal R}_0\delta_1\delta_m\times\frac{ \omega_{1a}}{\omega_1}\times
    \frac{\Lambda_{1a}}{\lambda_{1a}+\delta_{1a}}\times
       \\
&&\qquad \times    \left[1 +
      \frac{\delta_{1a}}{ 2(\lambda_{1a} +\delta_{pr\,a})}\right],
        \nonumber\\
\Lambda_1 &=&  \frac{V\left(\int f_0(\vec r_\bot) f_1(\vec r_\bot)
        u_{z} d\vec r_\bot\right)^2}{
    \int |f_0|^2 d\vec r_\bot \int |f_1|^2 d\vec r_\bot
    \int |\vec u|^2 d V },\nonumber\\
\Lambda_{1a} &=& \frac{V\left(\int f_0(\vec r_\bot) f_{1a}(\vec r_\bot)
        u_{ z} d\vec r_\bot\right)^2}{
  \int |f_0|^2 d\vec r_\bot \int |f_{1a}|^2 d\vec r_\bot
    \int |\vec u|^2 d V }\nonumber\,,\\
\lambda_{1}&=&\lambda +i\Delta\omega_{1}, \quad
    \Delta\omega_{1} =\omega_{0} -\omega_1-\omega_m\nonumber\,,\\
\lambda_{1a}&=&\lambda +i\Delta\omega_{1a}, \quad
    \Delta\omega_{1a} =\omega_{1a} -\omega_0-\omega_m\nonumber\,.
\end{eqnarray}
The equation (\ref{shchar2}) is the generalization of
(\ref{shchar}). Hereinafter subscript $_1$ corresponds to the
Stokes mode and subscript $_{1a}$ to the anti-Stokes mode.
$\Lambda_1,\ \Lambda_{1a}$ are the overlapping factors for the
Stokes and anti-Stokes modes correspondingly. $f_0$, $f_1$ and
$f_{1a}$ are the functions of the optical fields distribution in
the main, Stokes and anti-Stokes optical modes correspondingly
over the mirror surface. Vector $\vec u$ is the spatial vector of
displacements in the elastic mode, $u_z$ is the component of $\vec
u$ normal to the mirror surface, $\int d\vec r_\bot$ corresponds
to the integration over the mirror surface, and $\int dV$ over the
mirror volume $V$.

Analyzing  this characteristic equation one can see that the  presence
of the anti-Stokes mode can considerably depress or even 
exclude parametric instability. For example, let the main,  Stokes
and anti-Stokes modes be equidistant and belong to the main
frequency sequence $\omega_1 =\pi (K-1) c/L,\ \omega_0=\pi K c/L,
\ \omega_{1a}= \pi (K+1)c/L$ ($K$ is an integer).  In this case
$\Delta\omega_1=\Delta\omega_{1a}$, the main, Stokes and
anti-Stokes modes have the same gaussian distribution over the
cross section and hence the same overlapping factors:
$\Lambda_1=\Lambda_{1a}$. It means that the  second term in the
right part of (\ref{shchar2}) is larger than first term, the {\em
positive} damping introduced into elastic mode by the anti-Stokes
mode is greater than {\em negative} damping due to the Stokes
mode, hence the parametric instability is impossible. This case
has been analyzed in details in \cite{ak}.

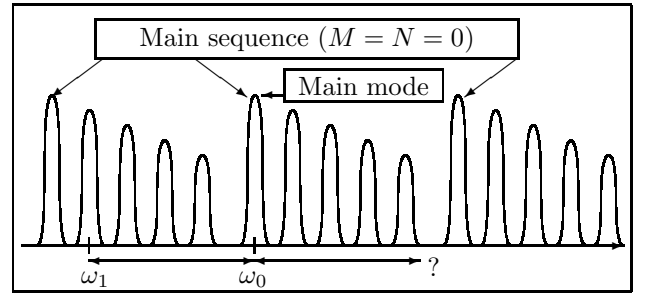
\begin{figure}
\unitlength0.05mm
\linethickness{0.5pt}
\begin{picture}(1642,762)
\bezier{220}(60,120)(80,120)(80,320)
\bezier{220}(80,320)(80,520)(100,520)
\bezier{220}(100,520)(120,520)(120,320)
\bezier{220}(120,320)(120,120)(140,120)
\bezier{220}(160,120)(180,120)(180,320)
\bezier{180}(180,320)(180,480)(200,480)
\bezier{180}(200,480)(220,480)(220,320)
\bezier{220}(220,320)(220,120)(240,120)
\bezier{180}(260,120)(280,120)(280,280)
\bezier{180}(280,280)(280,440)(300,440)
\bezier{180}(300,440)(320,440)(320,280)
\bezier{180}(320,280)(320,120)(340,120)
\bezier{160}(360,120)(380,120)(380,260)
\bezier{160}(380,260)(380,400)(400,400)
\bezier{180}(400,400)(420,400)(420,240)
\bezier{140}(420,240)(420,120)(440,120)
\put(20,120){\vector(1,0){1600}}
\bezier{140}(460,120)(480,120)(480,240)
\bezier{140}(480,240)(480,360)(500,360)
\bezier{140}(500,360)(520,360)(520,240)
\bezier{140}(520,240)(520,120)(540,120)
\put(200,140){\line(0,-1){40}}
\bezier{220}(1140,120)(1160,120)(1160,320)
\bezier{220}(1160,320)(1160,520)(1180,520)
\bezier{220}(1180,520)(1200,520)(1200,320)
\bezier{220}(1200,320)(1200,120)(1220,120)
\bezier{220}(1240,120)(1260,120)(1260,320)
\bezier{180}(1260,320)(1260,480)(1280,480)
\bezier{180}(1280,480)(1300,480)(1300,320)
\bezier{220}(1300,320)(1300,120)(1320,120)
\bezier{180}(1340,120)(1360,120)(1360,280)
\bezier{180}(1360,280)(1360,440)(1380,440)
\bezier{180}(1380,440)(1400,440)(1400,280)
\bezier{180}(1400,280)(1400,120)(1420,120)
\bezier{160}(1440,120)(1460,120)(1460,260)
\bezier{160}(1460,260)(1460,400)(1480,400)
\bezier{180}(1480,400)(1500,400)(1500,240)
\bezier{140}(1500,240)(1500,120)(1520,120)
\bezier{140}(1540,120)(1560,120)(1560,240)
\bezier{140}(1560,240)(1560,360)(1580,360)
\bezier{140}(1580,360)(1600,360)(1600,240)
\bezier{140}(1600,240)(1600,120)(1620,120)
\bezier{220}(600,120)(620,120)(620,320)
\bezier{220}(620,320)(620,520)(640,520)
\bezier{220}(640,520)(660,520)(660,320)
\bezier{220}(660,320)(660,120)(680,120)
\bezier{220}(700,120)(720,120)(720,320)
\bezier{180}(720,320)(720,480)(740,480)
\bezier{180}(740,480)(760,480)(760,320)
\bezier{220}(760,320)(760,120)(780,120)
\bezier{180}(800,120)(820,120)(820,280)
\bezier{180}(820,280)(820,440)(840,440)
\bezier{180}(840,440)(860,440)(860,280)
\bezier{180}(860,280)(860,120)(880,120)
\bezier{160}(900,120)(920,120)(920,260)
\bezier{160}(920,260)(920,400)(940,400)
\bezier{180}(940,400)(960,400)(960,240)
\bezier{140}(960,240)(960,120)(980,120)
\put(640,140){\line(0,-1){40}}
\bezier{140}(1000,120)(1020,120)(1020,240)
\bezier{140}(1020,240)(1020,360)(1040,360)
\bezier{140}(1040,360)(1060,360)(1060,240)
\bezier{140}(1060,240)(1060,120)(1080,120)
\put(400,80){\vector(-1,0){200}}
\put(400,80){\vector(1,0){240}}
\put(880,80){\vector(-1,0){240}}
\put(880,80){\vector(1,0){200}}
\put(420,100){\line(-1,-2){20}}
\put(880,100){\line(-1,-2){20}}
\put(720,520){\vector(-1,0){60}}
\put(1100,40){?}
\put(220,620){\framebox(1120,100)[]{Main sequence ($M=N=0$)}}
\put(721,505){\framebox(419,81)[]{Main mode}}
\put(1340,620){\vector(-4,-3){140}}
\put(240,620){\vector(-4,-3){140}}
\put(480,620){\vector(4,-3){140}}
\put(0,0){\framebox(1640,760)[]{}}
\put(600,20){$\omega_0$}
\put(180,20){$\omega_1$}
\end{picture}\\*[2mm]
\caption{Schematic structure of optical (Laguer-Gauss)
    modes in the FP cavity. The modes of the main frequencies sequence
    are shown by higher peaks. It is shown that  Stokes mode with frequiency 
	$\omega_1$
	may not have suitable  anti-Stokes mode (it is denoted by
	question-mark).}\label{fig2}
\end{figure}

However, it is worth  noting that this situation is possible only
for the small part of the total number of optical modes (see
fig.~\ref{fig2}). Indeed, resonance conditions $\omega_0 \simeq
\omega_1 +\omega_m$ can be fulfilled with a relatively high
probability for many of the optical Stokes and mirror elastic
modes combinations. If we assume the main optical mode to be
gaussian with the waist radius of the caustic $w_0$ (the optical
field amplitude distribution in the middle between the mirrors is
$\sim e^{-r^2/w_0^2}$)), then the Stokes and anti-Stokes modes are
described by generalized Laguerre functions (Gauss-Laguerre beams)
and the set of frequency distances  between the main and Stokes
(anti-Stokes) modes is determined by three integer numbers:
\begin{eqnarray}
\omega_0-\omega_1 &=& \frac{\pi c}{L}\left(K -
    \frac{2(2N+M)}{\pi}\, \arctan\frac{L\lambda_0}{2\pi w_0^2}\right)\simeq
    \nonumber\\
\label{DeltaOpt}
&\simeq& \left(2.4\, K -
     0.66\, N - 0.33\, M  \right)\times 10^5 \,
    \mbox{s}^{-1},\\
\omega_{1a}-\omega_0 &\simeq&  \left(2.4\, K_a +
     0.66\, N_a + 0.33\, M_a  \right)\times 10^5 \,
    \mbox{s}^{-1}.\nonumber
\end{eqnarray}
where $\lambda_0$ is the wavelength, $K=0\pm 1,\, \pm 2\dots $ is
the longitudinal index, $N=0,\, 1,\, 2\dots $, and $M=0,\, 1,\,
2\dots $ are the radial and angular indices, other numerical
parameters are given in Appendix~\ref{param}. We see that full
depression of parametric instability takes place  only if the
Stokes and anti-Stokes modes belong to the main sequence i.e.
$M=N=0$, when $\Delta\omega_1=\Delta\omega_{1a}$ and the Stokes
and anti-Stokes modes have equal spatial gaussian distribution.
Such modes obviously present the small part of the total optical
modes number. Indeed for $K=1,\ N=M=0$ we have  from
(\ref{DeltaOpt}) $\omega_0-\omega_1 \simeq 2.4\times 10^5\ {\rm
sec}^{-1}$. However, our numerical calculations show (see below)
that the lowest elastic mode has the frequency of about nine times
smaller: $\omega_{m\, \rm lowest}\simeq  0.28\times 10^5\ {\rm
sec}^{-1}$ and {\em only} within the range between $0.28\times
10^5\ {\rm sec}^{-1}$ and $ 1.6\times 10^5\ {\rm sec}^{-1}$ there
are more than $50$ (!) elastic modes and each of them has to be
carefully considered as possible candidate for parametric
instability.

For the case when the Stokes and anti-Stokes modes do not belong
to the main sequence (non zero numbers $N$ and $M$) the
frequencies of the suitable Stokes and anti-Stokes modes are not
equidistant from the main mode (i.e. $\Delta\omega_1\ne
\Delta\omega_{1a}$) and have different spatial distributions (i.e.
$\Lambda_1\ne \Lambda_{1a}$). Illustration of this is given in
fig.~\ref{fig2}. For the shown Stokes mode (left to the main mode)
there is no suitable anti-Stokes mode (it should be located right
to the main one). In this case one can use the approximate
condition for the parametric instability (see details of
approximations in Appendix \ref{app3}):
\begin{eqnarray}
\label{PICONDAS}
&&\frac{{\cal R}_0\Lambda_{1}}{ \left (
  1+\frac{\Delta\omega_{1}^2}{\delta_1^2} \right) }
    \times
      \frac{2+\frac{\delta_1}{\delta_{pr}}+
        \frac{\Delta\omega_{1}^2}{\delta_{pr}^2} }{
          2\left(1+\frac{\Delta\omega_{1}^2}{\delta_{pr}^2} \right) }-\\
&& \quad -\frac{{\cal R}_0\Lambda_{1a}}{ \left (
  1+\frac{\Delta\omega_{1a}^2}{\delta_{1a}^2} \right) }\,
    \frac{\omega_{1a}}{\omega_1} \times
      \frac{2+\frac{\delta_{1a}}{\delta_{pr\,a}}+
        \frac{\Delta\omega_{1a}^2}{\delta_{pr\,a}^2} }{
          2\left(1+\frac{\Delta\omega_{1a}^2}{\delta_{pr\,a}^2} \right)
          }>1 \nonumber\,.
\end{eqnarray}
Here the second term in the right part describes the influence of
the anti-Stokes mode.

%\newpage
The above consideration for lossless mirrors can be generalized
for mirrors with losses. Analyzing the more important case when
only the FP cavity mirrors have losses with loss coefficient $R\ll
T$ for each of them we can easily show all previous formulas to be
valid taking the following substitutions into account:
\begin{eqnarray}
\delta_1 &\Rightarrow & \delta_{1R}=\delta_1(1+\eta),\quad
    \eta =\frac{2R}{T}\ll 1 ,\\
\delta_{pr} &\Rightarrow & \delta_{pr\,R}=
        \left(\frac{T_{pr}}{4}+ \eta\right) \delta_1\,.
\end{eqnarray}
We see that in the case $\eta\ll T_{pr}/4$ all formulas does not
change, in the case of  $1\gg \eta\gg T_{pr}/4$ the condition of
the ultimate resonance changes but the condition of partial
resonance remains unchanged. For LIGO-II the losses are negligible
(it is planned that $R\simeq 5$~ppm,
    $\eta\simeq 2\times 10^{-3}\ll T_{pr}/4$).

The account of losses is important when we consider optical modes
with high indices $M,\ N$ because their diffractional losses
increases for higher indices $M,\ N$. We have numerically
calculated the equivalent loss coefficient $R_{M,\, N}$ describing
the diffractional losses on each mirror and have found that
\begin{eqnarray*}
%\label{dif1}
\eta_{\rm dif}=\frac{2R_{M,\,N}}{T}<\frac{T_{pr}}{4}\simeq 0.015,&\quad &
    \mbox{if}\ \ 2N+M\le 6,\\
\label{dif2}
\eta_{\rm dif}< 1,&\quad &
    \mbox{if}\ \ 2N+M\le 9,
\end{eqnarray*}
We see that there is a wide range of indices $M,\,N$ which must be
taken into account when analyzing parametric instability. Note
that even for the case $\eta_{\rm dif}>1$ the parametric
instability may also be possible due to the large numeric value of
factor ${\cal R}_0$.

Using the Femlab program package we have numerically calculated
the first 50 elastic modes frequencies and spatial distributions
in the cylindric mirror fabricated from fused silica (parameters
are listed in Appendix \ref{param}). These frequencies lied within
the range between $28000 \ {\rm sec}^{-1}$ and $164000 \ {\rm
sec}^{-1}$. We have estimated the errors $\Delta\omega_m$ of these
frequencies  comparing the data obtained with different total
number of nodes ($N_n\simeq 1135$ and $N_n\simeq 3 777$):
$\Delta\omega_m\simeq 500\, \cdots\, 4000\ {\rm sec}^{-1}$ (the
error increases for elastic modes at higher frequencies). Such
error is unacceptably large because  in order to determine the
ultimate resonance in power recycled interferometer the error have
to be $\Delta\omega_m \ll \delta_{pr}\simeq 1.5\ {\rm sec}^{-1}$
or in the case of "partial resonance" $\Delta\omega_m \ll
\delta_1\simeq 94\ {\rm sec}^{-1}$.

It is worth noting that the frequency density of elastic modes
rapidly increases at higher frequencies. In particular, the mean
distance $\Delta\omega_{md}$ between the elastic modes frequencies
can be estimated as follows:
$$
\Delta\omega_{md} \simeq \frac{2\omega_{m\, \rm lowest}^3}{\pi
\omega_m^2}\,,
$$
and it is equal to $\Delta\omega_{md}\simeq 100\ {\rm
sec}^{-1}\simeq\delta_1$ even at $\omega_m=3.8\times 10^5\ {\rm
sec}^{-1}$. It means that practically for each elastic mode with
frequency higher than $3.8\times 10^5\ {\rm sec}^{-1}$ there
exists the Stokes mode with small detuning: $\Delta\omega_1 <
\delta_1$ ("partial resonance"). However, the same speculations
can also be referred to the anti-Stokes modes and hence the
accurate calculation of overlapping factors $\Lambda_1,\
\Lambda_{1\,a}$ is required for each of the elastic modes.

Even in the case of parametric resonance the overlapping factor
$\Lambda_{1}$ may be zero (for example, elastic mode and the
Stokes mode can have different dependence on azimuth angle).
However, it is important to take into account that only the
elastic mode attached to the mirror axis in contrast to the
optical mode which can be shifted from the mirror axis due to
non-perfect optical alignment. Hence, the overlapping factor must
depend on distance $Z$ between the center of mirror and the center
of the main optical mode distribution over the mirror surface. It
means that $\Lambda_1$ may be zero for $Z=0$ but nonzero for $Z\ne
0$. Therefore, the numerical analysis of the mode structure should
evidently include the case when $Z\ne 0$. Note that there is a
proposal to use special shift $Z$ of the laser beam of about
several centimeters from the mirror axis in order to decrease
thermal suspension noise \cite{levin}.

Due to the necessity to decrease the level of thermoelastic and
thermorefractive noises \cite{bgv,bgvrefr,kip,thorneft} the size
of the light spot on the mirror surface is likely to be
substantially larger and the light density distribution in the
spot is not likely to be gaussian (the "mexican hat" modes
\cite{thorneft}) to evade  substantial diffractional losses. The
optical modes which are complementary to such a "mexican hat" main
mode have the frequencies more close to the main mode than given
by equation (\ref{DeltaOpt}), and the probability to be entrapped
into the parametric instability is higher. Thus the estimates
presented above  for gaussian optical modes may be regarded only
as the first approximation in which the use of analytical
calculations is still possible.

\section{Conclusion}

Summing up the above calculations and considerations we have to
conclude that the effect of parametric oscillatory instability is
a potential danger for the gravitational wave antennae with
powerful laser pumping. From our point of view, to estimate
correctly this danger it is necessary to implement the following
subprograms of researches:

\begin{enumerate}
\item To calculate numerically the values of eigenfrequencies $\omega_m$
for lower elastic modes with relative errors of at least $10^{-4}
\, \cdots 10^{-3}$. It is not an easy task because the error in
standard schemes of finite elements calculations rises as square
 of $\omega_m$. It is necessary to keep in mind that these calculations will
play a role of introductory ones because it is likely that in LIGO-II
non-gaussian mode distribution of light will be used (so called
"mexican hat" mode) and, correspondingly, it will be necessary to
calculate numerically all spectrum of "mexican hat" modes.

\item   At the same time the numerical analysis can not solve the problem
    completely
    because the fused silica pins and suspension fibers will be
    attached to the mirror. This attachment will change the elastic
    modes frequency values (and may be also the quality factor and distribution).
    For example, assuming the pin mass of about $\Delta m \simeq 80$~g one can estimate
    that frequency shift may be about
$$
\Delta \omega_{m\, pin}\le \pm
    \frac{\omega_m}{1\times 10^5\ {\rm sec}^{-1}}\times
    \frac{ \Delta m}{2m}\simeq \pm 100\ {\rm sec}^{-1},
$$
    i.e. about the value of $\delta_1$ (!).
    In addition, the unknown Young modulus and fused silica density
    inhomogeneity (we estimate it may be of about $10^{-3}\, \cdots \, 10^{-2}$)
    will additionally limit the numerical analysis accuracy. Thus
    we have to conclude that {\em the direct measurements}
    of eigenfrequency values, distribution and quality factors
    for several hundreds of elastic modes for each mirror
    of FP cavity are {\em inevitably} necessary.

\item
    When more "dangerous" candidates of elastic and Stokes modes will
    be known their undesirable influence can be possibly
    decreased. Perhaps,
    it can be done  by {\em accurate}
    small change of mirror shape or by introducing low noise damping
    \cite{bv}.

\item The last stage of this program  should be presented by the {\em direct tests} of the
    optical field behavior with smooth increase of the input optical power: it will
    be possible to register the appearance of the photons  at the Stokes
    modes and the rise of the $Q_m$ in the corresponding elastic mode
    while the power $W$ in the main optical mode is below the critical
    value.
\end{enumerate}
We think that the parametric oscillatory instability effect can be
overcome in the laser gravitational antennae after these detailed
investigations.

\section*{Acknowledgements}

This work was supported in part
by NSF and Caltech grant PHY0098715,
by Russian Ministry of Industry and Science
and by Russian Foundation of Basic Researches.

\appendix

\section{Power Recycled Interferometer}\label{app1}

In this Appendix we deduce the equations (\ref{D1a}, \ref{Xb}).

\subsection{Internal FP Cavities}

For the fourier components of complex slowly changing amplitudes
in general case we have obvious expressions (see notations in
fig.~\ref{fig1}):
\begin{eqnarray}
\label{F1ina}
F_1^{in}(\Omega)&=& \frac{2i\delta_1\, F_1(\Omega)}{\sqrt{T_1}\,
        (\delta_1 - i\Delta \omega -i\Omega)},\quad
    \delta_{1,2}= \frac {cT_{1,2}}{4L_{1,2}}, \\
\label{E1}
E_1(\Omega)     &=& - F_1(\Omega)\, \frac{\delta_1+i\Delta \omega+i\Omega}
        {\delta_1-i\Delta \omega-i\Omega}\,.
\end{eqnarray}
One can obtain the similar formulas for the second FP cavity:
\begin{eqnarray}
F_2^{in}(\Omega)&=& \frac{2i\delta_2\, F_2(\Omega)}{\sqrt{T_2}\,
        (\delta_2 - i\Omega)}, \\
\label{E2}
E_2(\Omega)     &=& - F_2(\Omega)\,
    \frac{\delta_2+i\Omega}
        {\delta_2-i\Omega}\,.
\end{eqnarray}

\subsection{The Mean and Small Amplitudes}

Let us introduce the mean amplitude and small amplitude (denoted
by small letters). For example, for amplitude in the FP I cavity
it means: $F_1^{in} = {\cal F}_1^{in}+f_1^{in}$.

Let us also assume that $\delta_2=\delta_1$, and now we keep in
mind that
\begin{equation}
\Delta\omega =\frac{\omega_1 x}{L_1}\,.
\end{equation}

Then the mean amplitudes are written as follows:
\begin{eqnarray}
\label{F1}
{\cal F}_1^{in} &=& \frac{2i}{\sqrt{T_1} }\, {\cal F}_1,\quad
    {\cal E}_1 = -{\cal F}_1,\\
\label{F2}
{\cal F}_2^{in} &=&  \frac{ 2i}{\sqrt{T_1} }\, {\cal F}_2,\quad
    {\cal E}_2 = -{\cal F}_2\,.
\end{eqnarray}
Rewriting the equations (\ref{E1} and \ref{E2}) we find the small
amplitudes:
\begin{eqnarray}
({\cal E}_1+e_1 )&\times &\big(\delta_1-i\Delta \omega-i\Omega\big)=\nonumber\\
&& -
    \big({\cal F}_1 +f_1\big)\, \big(\delta+i\Delta \omega+i\Omega\big),\\
\label{e1}
e_1 &=& -f_1\,\Gamma_0 -
    {\cal F}_1\,\Gamma_1,
\quad %\label{e2}
e_2 = -f_2 \Gamma_0\,, \\
\label{Gamma0}
\Gamma_0 &=& \frac{\delta_1+i\Omega}{\delta_1-i\Omega},\quad
    \Gamma_1=\frac{2i\Delta \omega}{\delta_1-i\Omega}\,.
\end{eqnarray}

\subsection{Beam Splitter}

Assuming that $T_{bs}=1/2$ one can write the following:

\begin{eqnarray*}
\label{F1a}
F_1 e^{-i\phi_1}    &=& \frac{1}{\sqrt{2}}\big( iF_3 + F_{4} \big),\quad
F_2 = \frac{1}{\sqrt{2}}\big( F_3 + iF_{4} \big),\\
E_4 &=& \frac{1}{\sqrt{2}}\big(iE_2 + E_1e^{i\phi_1} \big),\quad
E_3 =  \frac{1}{\sqrt{2}}\big(E_2 + iE_1e^{i\phi_1} \big)\,.
\end{eqnarray*}
Here $\phi_1=kl_1$ is the wave $E_1$ phase shift due to length
path $l_1$ between the FP cavity and the beam splitter. We assume
that $\phi_1=\pi/2$ and analogous phase shift $\phi_2$ (for the
wave $E_2$) is equal to zero. Then one can obtain for the mean
amplitudes the following:
\begin{eqnarray}
\label{F1m}
{\cal F}_{1} &=& \frac{-{\cal F}_{3}}{\sqrt 2 } ,\quad
    {\cal E}_{1} = \frac{{\cal F}_{3}}{\sqrt 2 },\\
{\cal F}_{2} &=& \frac{{\cal F}_{3}}{\sqrt 2 },\quad
    {\cal E}_{2} = \frac{-{\cal F}_{3}}{\sqrt 2 } \\
\label{E4m}
{\cal E}_4  &=& 0,\quad {\cal E}_3 =- {\cal F}_3,
\end{eqnarray}
Small amplitudes are equal to:
\begin{eqnarray}
\label{f1f}
f_1 &=& \frac{- f_{3}+if_4}{\sqrt 2 } ,\quad
    f_{2} = \frac{f_3 + i f_4}{\sqrt 2 },\\
\label{e1f}
e_4 &=& f_4\, \Gamma_0  +
    \frac{i{\cal F}_3 \Gamma_1 }{2},\quad
%\label{e3f}
e_3 =  -f_3\Gamma_0 - \frac{{\cal F}_3\Gamma_1}{2}.\nonumber
\end{eqnarray}

\subsection{Power Recycling Mirror}

We have the following expressions:
\begin{eqnarray}
F_3e^{-i\phi_{pr}} &=& i \sqrt{T_{pr}} F_5+\sqrt{1-T_{pr}} E_3 e^{i\phi_{pr}},\\
\label{E5}
E_5 &=&i\sqrt{T_{pr}} E_3 e^{i\phi_{pr}}+ \sqrt{1-T_{pr}} F_5,\\
\phi_{pr} &=& \frac{\big(\omega_0+\Delta\omega_{pr} +\Omega\big)l_{pr}}{c}.
\end{eqnarray}
Using (\ref{E4m}) and assuming that the PR cavity is in resonance:
$\exp(i\omega_0l_{pr}/c)= i $ (i.e. $\omega_0l_{pr}/c=\pi/2+2\pi
n$, $n$ is an integer) one can obtain:
\begin{eqnarray}
{\cal F}_3 &\simeq & \frac{-2{\cal F}_5}{\sqrt{T_{pr}} },\quad
    {\cal E}_3 \simeq  \frac{2{\cal F}_5}{\sqrt{T_{pr}} }\,,\\
\label{f3a}
f_3 A_3 &\simeq & -f_5\sqrt{T_{pr}}  +
    \frac{{\cal F}_3\Gamma_1 }{2},\\
\label{A3a}
A_3 &\simeq& 1-\Gamma_0 +   \Gamma_0 \frac{T_{pr}}{2}\left(1 -
        \frac{i\big(\Delta\omega_{pr} +\Omega\big)}{
            \tilde \delta_{pr}}\right)\,,\\
\tilde \delta_{pr} &=& \frac{cT_{pr}}{4L_{pr}}\,.\nonumber
\end{eqnarray}
The value of $\tilde \delta_{pr}\gg |\Delta\omega_{pr} +\Omega|$
and hence one can rewrite (\ref{A3a})  as follows:
\begin{eqnarray}
\label{A3b}
A_3 &\simeq & 1-\Gamma_0 +\Gamma_0  \frac{T_{pr}}{2}    =
2\, \frac{ \delta_{pr}- i\Omega}{\delta_1-i\Omega} ,\\
\delta_{pr}&=& \frac{T_{pr}\delta_1}{4}.
\end{eqnarray}
And finally, for $f_3$ one can obtain the following:
\begin{eqnarray}
\label{f3b}
f_3  &\simeq &\, \frac{\delta_1-i\Omega}{2(\delta_{pr}- i\Omega)}
    \left( -f_5\sqrt{T_{pr}}  +
    \frac{{\cal F}_3\Gamma_1 }{2}
    \right)\,.
\end{eqnarray}

Using (\ref{E5})we will have:
\begin{eqnarray}
\label{E5b}
{\cal E}_5 &=& -{\cal F}_5,\\
e_5 &=&  \sqrt{T_{pr}}\left(f_3\Gamma_0 +
    \frac{{\cal F}_3\, \Gamma_1}{2}\right) +f_5.
\end{eqnarray}

\subsection{The Small Variational Amplitude in the FP I Cavity}

In order to calculate the small amplitude $f_1$ one can rewrite
equation (\ref{F1ina}) using (\ref{f1f}) and (\ref{f3a}) assuming
that small amplitudes $f_4$ and $f_5$ describing vacuum
fluctuations of the input waves are zero:
\begin{eqnarray}
\left({\cal F}_1^{in}+f_1^{in}\right)&\times &
    \left(\delta_1 - i\Delta \omega -i\Omega\right)  =\\
&=& \frac{i\sqrt{2}\delta_1\,(-{\cal F}_3 - f_3 +i f_4)}{\sqrt{T_1}},
    \nonumber\\
\label{FF}
{\cal F}_1^{in} &=& \frac{2i{\cal F}_1}{\sqrt{T_1}}=
    -\frac{i\sqrt{2}\, {\cal F}_3}{\sqrt{T_1}}=
    \frac{2i\sqrt{2}\,{\cal F}_5}{\sqrt{T_1T_{pr}}},\\
f_1^{in} \left(\delta_1  -i\Omega\right) & = &
    i\Delta\omega\,{\cal F}_1^{in}      -   \\
&& -\frac{i\sqrt{2}\,\delta_1}{\sqrt{T_1}}
    \left(\frac{\delta_1-i\Omega}{2(\delta_{pr} - i\Omega)}\right)
    {\cal F}_3\, \frac{\Gamma_1}{ 2 }\,.
\end{eqnarray}
It can be rewritten using (\ref{FF}) as:
\begin{eqnarray}
\label{f1inb}
f_1^{in} \left(\delta_1  -i\Omega\right)(\delta_{pr} - i\Omega)
=   i{\cal F}_1^{in}\Delta\omega
    \left[ \delta_{pr} - i\Omega +  \frac{\delta_1}{2}\right]\,.
\end{eqnarray}

\subsection{Time Domain}

Now it seems that one can make in (\ref{f1inb}) the following
substitution:
\begin{eqnarray*}
{\cal F}_1^{in} &\to& D_0e^{-i(\omega_0-\omega_1)t}, \quad
\Delta\omega \to  \frac{\omega_1}{L_1}\,X^*e^{i\omega_m t}\,.
\end{eqnarray*}
However, such substitution will be  incorrect, because the
equation for the fourier transform of complex amplitude has not
contain time-dependent terms. The correct form of this equation is
the following:
\begin{eqnarray}
\label{f1inca}
f_1^{in}(\Omega) \left(\delta_1  -i\Omega\right)(\delta_{pr} - i\Omega)
= \\
\qquad  =
iD_0^{in}\, X^*(\Omega -\Delta\omega_{1})\frac{\omega_1 }{L_1}
    \left[\delta_{pr} - i\Omega +   \frac{\delta_1}{ 2}\right]\,.\nonumber
\end{eqnarray}

Now one can obtain time domain equivalent of (\ref{f1inca}) applying
inverse fourier transform:
\begin{eqnarray}
\label{f1inct}
 \left(\partial_t +\delta_1  \right)(\partial_t + \delta_{pr} )\, f_1^{in}(t)
&=&  \\
=iD_0^{in}\, e^{-i\Delta \omega_1 t}\,\frac{\omega_1 }{L_1}&\times &
    \left[\partial_t+ \delta_{pr} - i\Delta\omega_{1} +
        \frac{\delta_1}{ 2}\right]X^*(t)\nonumber =\\
= iD_0^{in}\, \,\frac{\omega_1 }{L_1} & \times &
\left[\partial_t+ \delta_{pr}  +
        \frac{\delta_1}{ 2}\right]X^*(t)\,e^{-i\Delta \omega_{1} t}\,.\nonumber
\end{eqnarray}
The last equation is the complex conjugate to (\ref{D1a}) where:
$$
f_1^{in}\to D_1.
$$

\subsection{Equation for Elastic Oscillations}

For elastic displacement we have the following equation:
\begin{eqnarray}
\partial_t^2 x &+& 2\delta_m \partial_t x +\omega_m^2 x =
    \frac{F_{pm}}{m},\\
F_{pm} &=& \frac{2S}{c}\, \frac{(E_0+E_1)^2}{4\pi}\simeq\nonumber\\
&\simeq&
    \frac{S}{\pi c}\left(
    {\cal F}_0^{in} (f_1^{in})^*e^{-i(\omega_0-\omega_1)t}+\right.\\
&& \qquad\left.+
    ({\cal F}_0^{in})^* f_1^{in}e^{i(\omega_0-\omega_1)t}
    \right), \nonumber
\end{eqnarray}
where $S$ is the cross section of light beam, $c$ is the light speed.
Introducing the slow amplitudes for displacement $x$:
$$
x(t) = X(t)\,e^{-i\omega_mt}+ X^*(t)\,e^{i\omega_mt}\,,
$$
one can obtain the following:
\begin{eqnarray}
\label{Xa}
\partial_t X +\delta_m X &=& \frac{iS}{2\pi c m \omega_m}
    {\cal F}_0^{in} (f_1^{in})^*e^{-i\Delta \omega_1 t},\\
\Delta \omega_1 &=& \omega_0-\omega_1-\omega_m.
\end{eqnarray}

This equation coincides with (\ref{Xb}) after the substitutions,
listed below:
$$
f_1^{in}\to D_1, \quad {\cal F}_0\to D_0\,.
$$

\section{Solution of the Characteristic Equation (\ref{PICOND})}\label{app2}

In this Appendix we obtain the instability condition
(\ref{PICOND}) from the characteristic equation (\ref{shchar}).

Let us write down the the solution of equation (\ref{shchar}) as
sum of real and imaginary part:
$$
\lambda=a+ib\,.
$$
The condition of instability is $a > 0$. Thus substituting
$\lambda = ib$ into (\ref{shchar}) one can find two equations
(introducing notations $A$ and $B$):
\begin{eqnarray}
\label{shcharr}
\delta_m &=&\underbrace{ {\cal R}_0\delta_1\delta_m\times
    \frac{\delta_1 }{(b+\Delta\omega_{1})^2+\delta_1^2}}_{A}\times\\
&& \times\underbrace{\left[1 +
      \frac{\delta_1\delta_{pr}-(b+\Delta\omega_{1})^2}{
      2[(b+\Delta\omega_{1})^2+\delta_{pr}^2]}\right]}_{B},\nonumber
\end{eqnarray}
\begin{eqnarray}
\label{shchari}
b &=& - {\cal R}_0\delta_1\delta_m\times
    \frac{ (b+\Delta\omega_{1})}{(b+\Delta\omega_{1})^2+\delta_1^2}
    \times \\
&& \times \underbrace{\left[1 +
      \frac{\delta_1(\delta_1+\delta_{pr})}{
      2[(b+\Delta\omega_{1})^2+\delta_{pr}^2]}\right]}_{C}=\nonumber \\
&=& - \frac{\delta_m (b+\Delta\omega_{1})C}{\delta_1B}\,.
\end{eqnarray}
Using the last equation  one can formally express $b$ as:
\begin{eqnarray}
\label{bb}
b &=& -\Delta\omega_1\,\left.\frac{\delta_m C}{\delta_1 B}\right/
  \left(1 +\frac{\delta_m C}{\delta_1 B} \right),\nonumber\\
\frac{C}{B} &=& \frac{
  2\delta_{pr}^2+\delta_1(\delta_1+\delta_{pr})+ 2(b+\Delta\omega_1)^2}{
    2\delta_{pr}^2+\delta_1\delta_{pr}+ (b+\Delta\omega_1)^2}\,.
\end{eqnarray}
It is obvious from the last expression that
$$
2 < \frac{C}{B} <\frac{\delta_1}{\delta_{pr} }
$$
for {\em any} value of $(b+\Delta\omega_1)$. Hence one can
conclude from (\ref{bb}) that $|b| \ll \Delta\omega_1$ (remind
that $\delta_m\ll \delta_1$) and equation (\ref{shcharr}) can be
simplified as follows:
\begin{eqnarray}
\delta_m &=& {\cal R}_0\delta_1\delta_m\times
    \frac{\delta_1 }{(\Delta\omega_{1}^2+\delta_1^2)}\times\\
&& \times
      \frac{\delta_{pr}(\delta_1+2\delta_{pr})+\Delta\omega_1^2}{
      2[\Delta\omega_{1}^2+\delta_{pr}^2]}\,.\nonumber
\end{eqnarray}
Now rewriting this equation one can easily obtain the condition of
parametric instability (\ref{PICOND}).

\section{Solution of the Characteristic Equation (\ref{PICONDAS}) }\label{app3}

In this Appendix using some approximation we deduce the
instability condition (\ref{PICONDAS}) from the characteristic
equation (\ref{shchar2}).

We write down the the solution of this equation as a sum of real
and imaginary parts anew:
$$ \lambda=a+ib $$
and assume  $a=0$.
So substituting $\lambda = ib$ into
(\ref{shchar2}) and extracting real and image parts we can write two equations:
\begin{eqnarray}
\label{charrAS}
\delta_m &=&\underbrace{ {\cal R}_0\delta_1\delta_m\times
    \frac{\delta_1\Lambda_{1}}{(b+\Delta\omega_{1})^2+\delta_1^2}}_{A_1}\times\\
&& \times\underbrace{\left[1 +
      \frac{\delta_1\delta_{pr}-(b+\Delta\omega_{1})^2}{
      2[(b+\Delta\omega_{1})^2+\delta_{pr}^2]}\right]}_{B_1} -\nonumber\\
& &- \underbrace{ {\cal R}_0\delta_1\delta_m\,\frac{\omega_{1a}}{\omega_1}\times
    \frac{\delta_{1a}\Lambda_{1a}}{
      (b+\Delta\omega_{1a})^2+\delta_{1a}^2}}_{A_{1a}}\times\nonumber\\
&& \times\underbrace{\left[1 +
      \frac{\delta_{1a}\delta_{pr\,a}-(b+\Delta\omega_{1a})^2}{
      2[(b+\Delta\omega_{1a})^2+\delta_{pr\,a}^2]}\right]}_{B_{1a}}=\nonumber\\
\label{deltam}
 &=& \underbrace{A_1B_1}_{\delta_{m\,1} }-
  \underbrace{A_{1a}B_{1a} }_{\delta_{m\,1a} }\,,
\end{eqnarray}
\begin{eqnarray}
\label{chariAS}
b &=& - {\cal R}_0\delta_1\delta_m\times
    \frac{\Lambda_{1}(b+\Delta\omega_{1})}{(b+\Delta\omega_{1})^2+\delta_1^2}
    \times \\
&& \times \underbrace{\left[1 +
      \frac{\delta_1(\delta_1+\delta_{pr})}{
      2[(b+\Delta\omega_{1})^2+\delta_{pr}^2]}\right]}_{C_1} + \nonumber \\
&&+{\cal R}_0\delta_1\delta_m\,\frac{\omega_{1a}}{\omega_1}\times
    \frac{\Lambda_{1a}(b+\Delta\omega_{1a})}{
      (b+\Delta\omega_{1a})^2+\delta_{1a}^2}\nonumber    \times \\
&& \times \underbrace{\left[1 +
      \frac{\delta_{1a}(\delta_{1a}+\delta_{pr\,a})}{
      2[(b+\Delta\omega_{1a})^2+\delta_{pr\,a}^2]}\right]}_{C_{1a}}=\nonumber
      \\ &=& - A_{1}C_{1}\, \frac{b+\Delta\omega_{1} }{\delta_{1} } +
  A_{1a}C_{1a}\, \frac{b+\Delta\omega_{1a} }{\delta_{1a} }\,.
\end{eqnarray}
Using notations (\ref{deltam}) this equation can be rewritten as:
\begin{eqnarray*}
\label{b}
b &=& -\frac{\delta_{m\,1} C_1 }{\delta_1B_1}\times (b+\Delta\omega_{1})+
  \frac{\delta_{m\,1a} C_{1a} }{\delta_{1a}B_{1a} }\times
  (b+\Delta\omega_{1a})\,,
\end{eqnarray*}
and one can formally express $b$ as:
\begin{eqnarray}
\label{bbb}
b &=& \frac{\left(-\Delta\omega_1\,\frac{\delta_{m1} C_1}{\delta_1 B_1}+
    \Delta\omega_{1\,a}\,
    \frac{\delta_{m1\,a} C_{1\,a} }{\delta_{1\,a} B_{1\,a} }\right)}
  {\left(1 +\frac{\delta_{m1} C_1}{\delta_1 B_1}-
    \frac{\delta_{m1\,a} C_{1\,a}}{\delta_{1\,1} B_{1\,a}} \right)}.
\end{eqnarray}
Using definitions (\ref{charrAS}, \ref{chariAS}) it is easy to prove that
\begin{eqnarray*}
\label{CB}
 2 <  \frac{C_1}{B_1}  <\frac{\delta_1}{\delta_{pr} } &\quad \mbox{or}\quad &
     \frac{2\delta_{m1} }{\delta_1} <  \frac{\delta_{m1} C_1}{\delta_1 B_1}
      <\frac{\delta_{m1} }{\delta_{pr} }\,,\\
 2 <  \frac{C_{1a} }{B_{1a} }  <\frac{\delta_{1a} }{\delta_{pr\,a} }
     &\quad \mbox{or}\quad &
     \frac{2\delta_{m1\,a} }{\delta_{\,a} 1} <
        \frac{\delta_{m1\,a} C_{1\,a}}{\delta_{1\,a} B_{1\,a}}
      <\frac{\delta_{m1\,a} }{\delta_{pr\,a} }.
\end{eqnarray*}
Now assuming that
\begin{equation}
\label{deltam1}
\frac{\delta_{m\,1} }{\delta_{pr}}\ll 1,\qquad
  \frac{\delta_{m\,1a} }{\delta_{pr\,a} }\ll 1\,,
\end{equation}
we can conclude that
\begin{equation}
\label{bsmall}
|b| \ll \Delta\omega_1,\ \Delta\omega_{1a}.
\end{equation}
Then the parametric instability condition (\ref{PICONDAS}) can be
easily obtained from (\ref{charrAS}).

It is worth  noting that inequalities (\ref{deltam1}) are the key
assumption for deduction of the PI condition (\ref{PICONDAS}).
These inequalities  obviously correspond to the case when values
$\delta_{m1}=A_1B_1$ and $\delta_{m1\,a}= A_{1a}B_{1a}$ are not
very close to each other. In order to understand how close they
can be  let us assume that
$$
\delta_{m1\,a}=\delta_{m1}(1-\epsilon),\quad \epsilon\ll 1,
$$
and try to estimate theminimal value of $\epsilon$. From
(\ref{deltam1}) one can obtain $\delta_{m1}\simeq
\delta_{m1\,a}\simeq \delta_m/\epsilon$. Hence the inequalities
(\ref{deltam1}) are equivalent to:
\begin{eqnarray}
\frac{\delta_m}{\epsilon\delta_{pr}}\ll 1\,.
\end{eqnarray}
Using parameters of LIGO-II (see Appendix \ref{param}) we have
estimates $\delta_m \simeq 1\times 10^{-3}\ {\rm sec}^{-1}$,
$\delta_{pr} \simeq 1\ {\rm sec}^{-1}$. Hence one can conclude
that for values of $\epsilon \ge 10^{-2}$ the inequalities
(\ref{deltam1}) fulfill.

\section{Numerical Parameters}\label{param}

For interferometer we use parameters planned for LIGO-II.
More details see in \cite{ligo2}.

\begin{equation}
    \begin{array}{lcllcl}
\omega_0 &=& 2\times 10^{15}\ {\rm sec}^{-1}, & w_0 &=& 5.5\ {\rm cm},\\
T &=& 5\times 10^{-3}, & L &=&4\times 10^5\ {\rm cm},\\
T_{pr}&=& 6\times 10^{-2},& l_{pr} &\simeq &  10 \ {\rm m},\\
\delta_1 & \simeq & 94\ {\rm sec}^{-1},  &
    \delta_{pr} &\simeq &  1.5\ {\rm sec}^{-1},\\
W &=& 830\ {\rm kW},  &
    {\cal E}_0 &\simeq & 2.2\times 10^8\  {\rm erg}
    \end{array}
\end{equation}
Here $W$ is the power circulating inside FP cavity (${\cal E}_0 =
\frac{2L\,W}{c}$). We assume that cylindric mirror (with radius
$R$, height $H$ and mass $m$) is fabricated from fused silica with
angle of structural losses $\phi= 1.2\times 10^{-8}$\cite{ageev}:
\begin{equation}
\begin{array}{lcllcl}
R &=& 19.4 \ {\rm cm}, & H &=&15.4 \ {\rm cm},\\
m &=& 40\ {\rm kg},& \rho&=& 2.2 \ {\rm g/cm}^3,\\
E &=& 7.2\times 10^{11}\,
    \frac{\mbox{erg}}{\mbox{cm}^3},  &
    \sigma &=& 0.17,\\
\phi &=& 1.2\times 10^{-8}, &
    \delta_m &=& \omega_m\phi/2.
\end{array}
\end{equation}
Here $E$ is Young's modulus, $\sigma$ is Poison ratio.
It is useful to calculate factor ${\cal R}_0$ for these parameters and
elastic mode frequency $\omega_m=1\times 10^5\ {\rm sec}^{-1}$:
\begin{eqnarray}
{\cal R}_0 & = & \frac{{\cal E}_0 }{2mL^2\omega_m^2}\,
    \frac{\omega_1\omega_m}{\delta_1\delta_m}=
    \frac{{\cal E}_0 \omega_1}{mL^2\omega_m^2\phi\,\delta_1}\,
            \simeq 6\,100 \nonumber
\end{eqnarray}


\begin{thebibliography}{99}
\bibitem{abr1} A.\ Abramovici {\it et al.}, {\em Science} {\bf 256}(1992)325.
\bibitem {abr2} A.\ Abramovici {\it et al.},
    {\em Physics Letters} {\bf A218}, 157 (1996).
\bibitem{amaldi} Advanced LIGO System Design  (LIGO-T010075-00-D),
    Advanced LIGO System requirements  (LIGO-G010242-00), available in
    {\sf http://www.ligo.caltech.edu }.
\bibitem{bsv} V.\ B.\ Braginsky, S.\ E.\ Strigin, and S.\ P.\ Vyatchanin,
    {\em Physics Letters} {\bf A287}, 331 (2001); gr-qc/0107079;
\bibitem {ak} E. D'Ambrosio and W. Kells, to be published in
    {\em Physics Letter A}. LIGO-T020008-00-D, available in
    {\sf http://www.ligo.caltech.edu }.
%\bibitem{french} M.Pinard, Z. Hadjar and A. Heidman, {\sf http//:xxx.lanl.gov/quant-ph/9901057} (1999).
%\bibitem{BM1} V. B. Braginsky and I. I. Minakova, Bull. of MSU, ser. III, no.1,        83 (1964).
%\bibitem{BM2} V. B. Braginsky A. B. Manukin, and M. Yu. Tikhonov,
%    Sov. Phys. JETP {\bf 58}, 1550 (1970).

\bibitem{ligo2} {\sf http://www.ligo.caltech.edu/$\sim$ligo2/scripts/12refdes.htm}
\bibitem{thorneft} V.\ B.\ Braginsky, E.\ d'Ambrosio, R.\ O'Shaughnessy,
    S.\ E.\ Strigin, K.\ Thorne, and S.\ P.\ Vyatchanin,
    reports on LSC Meetings: Baton Rouge, LA, 16 March 2001 and Hanford, WA,
    15 August 2001:
    LIGO documents G010151-00-R and G010333-00-D
    ({\sf http//:www.ligo.caltech.edu/}).
%\bibitem{mason} "Physical acoustics.  Principles and Methods."       Edited by W. P. Mason, Vol. I, {\em Methods and devices},Part A,    {\em Academic press} ,  New York and London (1964).
\bibitem{ageev} A. Yu. Ageev, S. D. Penn, private communications.
\bibitem{levin} V. B. Braginsky, Yu. Levin and S. P. Vyatchanin,
    {\em Meas. Sci. Technol} {\bf 10} (1999), 598-606.
\bibitem{bgv} V.\ B.\ Braginsky, M.\ L.\ Gorodetsky, and S.\ P.\ Vyatchanin,
    {\em Physics Letters} {\bf A264}, 1 (1999); cond-mat/9912139;
\bibitem{kip} Yu. T. Liu and K. S. Thorne, submit to {\em Phys. Rev. D}.
\bibitem{bgvrefr} V.\ B.\ Braginsky, M.\ L.\ Gorodetsky, and
    S.\ P.\ Vyatchanin, {\em Physics Letters},  {\bf A 271}, 303-307 (2000).
\bibitem{bv} V. B. Braginsky and S. P. Vyatchanin,
     {\em Physics Letters A}, {\bf 293}  (2002) 228-234.


\end{thebibliography}
\end{document}